\documentclass[twoside,12pt]{article}
\usepackage{CJK}
\usepackage{indentfirst}
\usepackage{bm}
\usepackage{epsfig}
\usepackage{graphicx}

\usepackage{amssymb}
\usepackage[pdfstartview=FitH, CJKbookmarks=true, bookmarksnumbered=true, bookmarksopen=true, colorlinks=true, pdfborder=0 0 1, citecolor=blue, linkcolor=blue, linktocpage=true] {hyperref}
\usepackage{natbib}

\begin{document}

\thispagestyle{empty} \vspace*{0.8cm}\hbox
to\textwidth{\vbox{\hfill\huge\sf Commun. Theor. Phys.\hfill}}
\par\noindent\rule[3mm]{\textwidth}{0.2pt}\hspace*{-\textwidth}\noindent
\rule[2.5mm]{\textwidth}{0.2pt}


\begin{center}
\LARGE\bf Real-space construction and selection rules of center-of-mass momentum eigenstates in two-dimensional many-body lattices$^{*}$
\end{center}

\footnotetext{\hspace*{-.45cm}\footnotesize $^\dag$Corresponding author, E-mail: chleecn@szu.edu.cn}

\begin{center}
\rm Ling Lin$^{\rm{a,b})}$, \ \ Yongguan Ke$^{\rm{a})}$, \ and  \ Chaohong Lee$^{\rm{a,b})\dagger}$
\end{center}

\begin{center}
\begin{footnotesize} \sl
{Institute of Quantum Precision Measurement, State Key Laboratory of Radio Frequency Heterogeneous Integration, College of Physics and Optoelectronic Engineering, Shenzhen University, Shenzhen 518060, China${}^{\rm a)}$ \\
 Quantum Science Center of Guangdong-Hong Kong-Macao Greater Bay Area (Guangdong), Shenzhen 518045, China${}^{\rm b)}$ \\
}
\end{footnotesize}
\end{center}

\begin{center}
\footnotesize (Received XXXX; revised manuscript received XXXX)

\end{center}

\vspace*{2mm}

\begin{center}
\begin{minipage}{15.5cm}
\parindent 20pt\footnotesize
We present a comprehensive framework for the real-space construction of center-of-mass (c.m.) momentum eigenstates in two-dimensional (2D) quantum many-body systems.
Unlike standard momentum-space approaches, our real-space formalism explicitly captures the non-uniformity of cyclic subspace dimensions, which are fundamentally governed by the spatial symmetries of the many-body configurations and particle statistics.
We demonstrate that performing 2D real-space translations on multi-particle states naturally induces non-trivial translation phase factors associated with the characters of these stabilizer subgroups, leading to a renormalization of permissible momentum values. 
Notably, we identify an anomalous cross-dimensional momentum coupling in highly symmetric particle distributions, where the permissible momentum values in one direction are strictly constrained by those in the perpendicular direction. 
We emphasize the algorithmic necessity of this coupling, as neglecting it leads to missing or duplicated basis states in numerical implementations.
Concrete examples are provided to demonstrate the practical implementation and physical consequences of this framework.
\end{minipage}
\end{center}

\begin{center}
\begin{minipage}{15.5cm}
\begin{minipage}[t]{2.3cm}{\bf Keywords:}\end{minipage}
\begin{minipage}[t]{13.1cm}
Translation symmetry, center-of-mass momentum, many-body physics
\end{minipage}\par\vglue8pt

\end{minipage}
\end{center}

\section{Introduction}
%
%
The many-body Hamiltonian can be separated into different independent sectors labelled by the center-of-mass (c.m.) momentum (also called the total momentum) by ensuring the translation symmetry in periodic lattices.
This block-diagonalized structure not only simplifies the derivation of multi-particle band theories and excitation spectra but also enables efficient numerical simulations of large-scale systems.
Consequently, exploiting translational symmetry is beneficial for uncovering the rich physics of lattice systems in many-body quantum physics, especially for studying interacting topological states \cite{PhysRevB.87.205136,Topo_flat_S021797921330017X,PhysRevX.1.021014}.
Exact diagonalization (ED) of many-body quantum systems with translation symmetry typically relies on constructing c.m. momentum basis states, which is the eigenstate of multi-particle translation operator.
As long as the translation symmetry is preserved, bases with different c.m. momenta will not be coupled by the Hamiltonian.
In general, there are two approaches to construct the c.m. momentum basis, namely the momentum-space method and the real-space method.
The former implements under the Hilbert space spanned by the Fock basis in momentum space $\{ |{{\bf{k}}_1},{{\bf{k}}_2}, \cdots ,{{\bf{k}}_N}\rangle \} $ where ${\bf{k}}_j$ denotes the single-particle momentum of $j$-th particle.
These bases are naturally eigenstates of the multi-particle translation operator.
However, it renders particle-particle interactions highly non-local, resulting in dense Hamiltonian matrices that are computationally challenging for large systems.
On the other hand, extracting real-space physical observables, such as density distributions, requires cumbersome transformations. 
In contrast, the real-space method, detailed below, operates directly on position basis $\{ |{{\bf{r}}_1},{{\bf{r}}_2}, \cdots ,{{\bf{r}}_N}\rangle \} $ with ${\bf{r}}_j$ being the position of $j$-th particle.
While less intuitive for momentum conservation, it offers distinct advantages: it usually preserves the sparsity of the Hamiltonian, as interactions become diagonal and the tunneling is usually finite-range in many systems.
Moreover, real-space construction facilitates the study of systems with constrained kinetic terms, such as those encountered in many-body localization \cite{PhysRevLett.124.160604, PhysRevLett.130.120401, PhysRevB.109.184313}. 
In these cases, the complexity of the kinetic operator often precludes a simple transformation to the momentum-space representation.
Consequently, real-space formulations become essential for capturing the correct dynamics.
Despite these advantages, constructing c.m. momentum eigenstates in real space for multi-particle systems poses challenges different from single-particle scenarios. 
The procedure usually proceeds in two steps \cite{BARWINKEL2000240, Sandvik2010, jung2020guide,PhysRevB.107.125161}: first, we enumerate all cyclic subspaces; second, c.m. momentum eigenstates are formed in a Bloch-wave-like fashion, which involves translating a specific multi-particle position basis (a ``seed" state).
In constructing c.m. momentum states, one may find that certain momentum values are forbidden for some invariant subspaces of the co-translation operator.
Therefore, the dimension of the cyclic subspace may vary significantly.
Meanwhile, different particle statistics may also lead to different situations of the permissible momentum value.
Fully understanding the reason of this phenomenon is needed.

 In this article, we focus on the real-space construction of c.m. momentum basis of multi-particle systems in 2D lattice.
While real-space momentum construction principles have been discussed in the literature~\cite{Sandvik2010, jung2020guide, PhysRevB.107.125161}, a systematic and operational formulation tailored for complex 2D many-body configurations remains highly desirable. 
To prevent potential algorithmic pitfalls and clarify the underlying group-theoretical mechanisms, this work advances the current real-space exact diagonalization (ED) framework with three distinct contributions:
\begin{itemize}
    \item[\textbf{1.}] \textbf{Unified Construction Workflow:} We establish a rigorous and unified real-space construction protocol for 2D multi-particle center-of-mass (c.m.) momentum bases, clarifying how the stabilizer subgroup structure of the seed configuration orbits fundamentally dictates the varying dimensions of cyclic subspaces.
    \item[\textbf{2.}] \textbf{Operational Selection Rules:} We systematically classify the non-trivial translation phase factors arising from the interplay between quantum statistics and high spatial symmetries, deriving explicit, ready-to-code momentum selection rules.
    \item[\textbf{3.}] \textbf{Orthogonal Momentum Coupling:} Through concrete 2D lattice examples, we unveil that the permissible eigenvalues of $K_y$ can analytically depend on the choice of $K_x$. We elucidate the algorithmic consequence of this coupling, demonstrating that improper handling of this geometric entanglement inevitably leads to severe errors such as missing or duplicated basis states in ED programs.
\end{itemize}

The rest of this paper is structured as follows: In Sec.~\ref{sec:Momentum_space_Method}, we briefly review the conventional momentum-space method for constructing center-of-mass momentum basis states.
In Sec. \ref{sec:Real_space_Method}, we detail our real-space construction framework.
We systematically introduce the method for generating the momentum basis along a single direction, extend it to two-dimensional systems to reveal the non-trivial orthogonal momentum coupling, and outline the procedure for efficiently evaluating the Bloch Hamiltonian in real space. 
In Sec.~\ref{sec:numerical_computation}, we present numerical computations to compare the momentum-space and real-space methods.
Finally, we summarize our main findings and discuss their broader physical implications in Sec.~\ref{sec:summary_and_discussion}.

\section{Momentum-space Method}
\label{sec:Momentum_space_Method}
Firstly, let us give a brief introduction to obtain the c.m. momentum states in the momentum-space representation.
Without loss of generality, we consider the periodic boundary condition (PBC) in a 2D lattice system and restrict ourselves to the translation-invariant condition.
Meanwhile, we only consider two-body interactions between particles throughout this work.
For simplicity, we assume a $\mathrm{U}(1)$ symmetry for the corresponding Hamiltonian.
Generally, it is quite easy to diagonalize the kinetic part of the second-quantized Hamiltonian in the momentum space for bosons or fermions:
\begin{eqnarray}
\label{eqn:Ham_momentum_representaion}
\hat H &=& \sum\limits_{n, \boldsymbol{k}} {{\varepsilon _{n,{\boldsymbol{k}}}}\hat \gamma _{n,{\boldsymbol{k}}}^\dag {{\hat \gamma }_{n,{\boldsymbol{k}}}}}  \nonumber \\
 &&+ \sum\limits_{\{ n,{\boldsymbol{k}}\} } {V_{{{\boldsymbol{k}}_1},{{\boldsymbol{k}}_2},{{\boldsymbol{k}}_3},{{\boldsymbol{k}}_4}}^{{n_1},{n_2},{n_3},{n_4}}\hat \gamma _{{n_1},{{\boldsymbol{k}}_1}}^\dag \hat \gamma _{{n_2},{{\boldsymbol{k}}_2}}^\dag {{\hat \gamma }_{{n_3},{{\boldsymbol{k}}_3}}}{{\hat \gamma }_{{n_4},{{\boldsymbol{k}}_4}}}} ,
\end{eqnarray}
where $\hat \gamma _{n,{\boldsymbol{k}}}^\dag  =\sum_\sigma {u _{\sigma ,{\boldsymbol{k}}}}\hat c_{\sigma ,{\boldsymbol{k}}}^\dag $, with ${\hat c_{\sigma, {{\boldsymbol{k}}}}^\dag }$ being the creation operator of the particle with momentum $\boldsymbol{k}$, and $\sigma$ denotes the internal degree of freedom within the unit cell.
$\varepsilon _{n,{\boldsymbol{k}}}$ is the eigenenergy of the kinetic term, and the last term of Eq.~\ref{eqn:Ham_momentum_representaion} corresponds to the particle-particle interaction part with momentum-dependent interaction strength $V_{ \{ \boldsymbol{k_1} , \cdots \} }^{ \{ n_1, \cdots \}}$, describing a pair-wise elastic scattering among particles.
We assume that the particle-particle interaction strength only depends on relative distance so that the translation symmetry is always preserved.
Thus, the single-particle momenta satisfy the momentum conservation constraint ${\boldsymbol{k}}_1 + {\boldsymbol{k}}_2 = {\boldsymbol{k}}_3 + {\boldsymbol{k}}_4 \; \mathrm{mod} \; \boldsymbol{G}$, where $\boldsymbol{G}$ denotes the reciprocal lattice vector.
Notably, the momentum representation of the hard-core boson is complicated.
To properly study the hard-core boson in momentum space, we need to perform the Jordan-Wigner transformation to obtain the equivalent fermionic representation.
For a translationally invariant system, where the particle-particle interaction only depends on the relative distance, the c.m. momentum (total momentum) $\boldsymbol{K} = \sum_{j} {\boldsymbol{k}_{ j}}$ is a good quantum number.
Then, the $N$-particle momentum basis can be expressed as
\begin{equation}
\label{eqn:momentum_basis_momentum_rep}
|{\boldsymbol{K}},\alpha \rangle  \equiv \hat{\mathcal{T}}{\prod\limits_{j=1}^{N} { {\hat \gamma _{{n_j},{{\boldsymbol{k}}_j}}^\dag } } }|0\rangle ,
\end{equation}
where $\alpha$ is used to distinguish different relative relations among the single-particle momenta $\{ \boldsymbol{k}_j \}$ and $\hat{\mathcal{T}}$ is the ordering operator.
This set of basis can be generated efficiently in numerical computation \cite{Zhang_2010}.
We have assumed an ascending order of single-particle momenta in Eq.~\ref{eqn:momentum_basis_momentum_rep}. 
In numerical computations, this ordering is strictly enforced by appropriately permuting the particle operators, thereby naturally accounting for particle statistics.
The preservation of translation symmetry implies that the subspaces corresponding to specific center-of-mass momenta are invariant under the action of the Hamiltonian. 
Mathematically, this allows the Hamiltonian matrix to be reduced to a block-diagonal structure, such that $\langle {\bf{K}},\alpha |\hat H|{\bf{K}}',\alpha '\rangle  = 0$ for all ${\bf{K}} \neq {\bf{K}}'$. 
Thus, the eigenvalue problem can be solved independently within each momentum sector.
As the kinetic part is diagonal, this method allows one to perform the band projection on to certain gapped single-particle bands for multi-particle systems \cite{PhysRevX.1.021014, Topo_flat_S021797921330017X, PhysRevResearch.5.013112} to simplify the computation.
Provided that the lowest band gap of the kinetic part is large enough compared with the inter-particle interaction strength, one can neglect other bands.
For example, if we project the system onto the lowest gapped band $n_j=0$, there is ${\hat \gamma _{{n_j},{{\boldsymbol{k}}_j}}^\dag } = 0$ if $n_j>0$.
The band projection will reduce the Hilbert space dimension and thus benefit the numerical computation.
However, although the kinetic term is diagonal, evaluating the interaction terms remains challenging. 
This stems from the fact that inter-particle interactions couple different single-particle momentum states, rendering them highly non-local in c.m. momentum space.
Moreover, the Bloch Hamiltonian ${H\left( {\boldsymbol{K}} \right)} $ generally loses its sparseness in this momentum-space representation, bringing trouble to the large-scale numerical diagonalization algorithm, e.g. the Lanczos method.
This may cost more computation resources.
Moreover, under the momentum representation, the real-space particle distribution is not obvious.
It usually requires considerable efforts to transform the momentum representation back to the real-space representation and then compute the real-space density distribution.
Hence, working under the real-space representation to construct the momentum basis can be beneficial in some cases.

\section{Real-space Method}
\label{sec:Real_space_Method}
Next, we introduce the method to construct the momentum basis in real space (position space), which has some advantages compared to the momentum-space method.
%
The real-space method can be directly applied to hard-core bosons because the real-space occupation basis naturally enforces the hard-core constraint without requiring an explicit and complex momentum-space Jordan-Wigner transformation in two dimensions.
It provides a direct approach for studying the feature of relative distributions between particles, providing a potential route to study some specific phenomenon such as the fractional topological insulator.
Meanwhile, the real-space method also circumvents the calculation of complex interaction vertices in momentum space, enabling some more complex interaction forms.
Here,  we focus on a 2D case, where the construction of momentum basis in real space is much more complicated.
The multi-particle basis in position space can be written as
\begin{equation}
|{{\bf{r}}_{1}},{{\bf{r}}_{2}}, \cdots ,{{\bf{r}}_{N}}\rangle  \equiv \hat {\cal T}\prod\limits_{j = 1}^N {\hat c_{{{\bf{r}}_j}}^\dag } |0\rangle ,
\end{equation}
where ${\hat c_{{{\boldsymbol{r}}_j}}^\dag }$ is the bosonic or fermionic creation operator of the $j$th particle positioned at $\boldsymbol{r}_j$.
This set of real-space basis can be also generated efficiently \cite{Zhang_2010}.
Our main idea is to construct the eigenstate of the translation operator $\hat{T}_{\mu}, \mu= x, y$ through these real-space bases.
The translation operator shifts all particles by one unit cell
\begin{equation}
{{\hat T}_\mu }\hat c_{{{\bf{r}}_j}}^\dag {{\hat T}_\mu }^{ - 1} = \hat c_{{{\bf{r}}_j} + {{\boldsymbol{a}}_\mu }}^\dag ,
\end{equation}
and acts on multi-particle basis as
\begin{equation}
{{\hat T}_\mu }|{{\bf{r}}_1}, \cdots ,{{\bf{r}}_N}\rangle  = |{{\bf{r}}_1} + {{\boldsymbol{a}}_\mu }, \cdots ,{{\bf{r}}_N} + {{\boldsymbol{a}}_\mu }\rangle ,
\end{equation}
with $\boldsymbol{a}_{\mu}$ being the primitive lattice vector along $\mu=x,y$ direction.
Under PBC, the translation operator satisfies 
\begin{equation}
\langle {{\bf{r}}_1}, \cdots ,{{\bf{r}}_N}|{\left( {{{\hat T}_\mu }} \right)^{{L_\mu }}}|{{\bf{r}}_1}, \cdots ,{{\bf{r}}_N}\rangle  = 1,
\end{equation}
in which $L_\mu$ is the number of cells along $\mu$th direction.
By applying the translation operator for at most $L_\mu$ times, we can find a set of real-space states spanning a invariant subspace of the translation operator
\begin{equation}
V_\beta = \left\{ {\left. {{{\left( {{{\hat T}_x}} \right)}^m}{{\left( {{{\hat T}_y}} \right)}^l}|{\rm{seed}},\beta \rangle } \right|m,l \in \mathbb{Z} } \right\},
\end{equation}
in which $|{\rm{seed}},\beta \rangle  \equiv |{{\bf{r}}_1}, \cdots ,{{\bf{r}}_N}\rangle $ denotes some certain particle distributions and we call it the \emph{seed basis}.
The seed basis corresponds to one of the vector in its cyclic subspace, and $\beta$ is the index characterizing different seed bases.
We treat the translation group $G=\mathbb{Z}_{L_x} \times \mathbb{Z}_{L_y}$ as acting on the set of all $N$-particle configurations. 
This action partitions the Hilbert space into disjoint ``orbits", each generated by a representative seed basis. 
Without loss of generality, we have assumed that translation operations along different direction commute with each other: $\left[ {{{\hat T}_\mu },{{\hat T}_{\mu '}}} \right] = 0$.
Through this method, one can find all different cyclic subspaces of the translation operator iteratively.
The full Hilbert space is the sum of these cyclic subspaces $\mathcal{H} = \oplus_{\beta} {V_\beta}  $.
%
%
Then, the eigenstate of the translation operator can be written as the following plane-wave expression \cite{10.1063/1.168440} based on the seed basis:
\begin{equation}
\label{eqn:K_basis_direct}
|{\bf{K}},\beta \rangle  = \frac{1}{{\sqrt {{{\cal N}_\beta }} }}\sum\limits_{m,l} {{e^{i{K_x}m}}{e^{i{K_y}l}}{{\hat T}_x}^m{{\hat T}_y}^l|{\rm{seed}},\beta \rangle } ,
\end{equation}
in which ${{{\cal N}_\beta }}$ is the normalization factor.
It should be noted that, given a seed basis, permissible values of momentum is determined by the dimension of corresponding cyclic subspace, which is $\mathrm{dim}(V_\beta) = {{{\cal N}_\beta }}$.
%
%
The construction of the c.m. momentum state from a specific seed basis is mathematically grounded in the orbit-stabilizer theorem. The allowed momentum sectors are determined by the characters of the little group (stabilizer subgroup) associated with each many-body configuration orbit~\cite{tinkham2003group, procesi2007lie}.
In practical calculations, we find it difficult to directly construct $|{\bf{K}},\beta \rangle$ using Eq.~\ref{eqn:K_basis_direct}.
Usually, ${{\cal N}_\beta }$ is equal to the number of total cells.
However, for some certain cases, where the particle distribution of seed basis is highly symmetric, the dimension of cyclic subspace can be smaller.
It means that it only takes fewer translation operations to return to the original seed basis.
Thus, permissible values of momenta are restricted.
The simplest example is to consider a extremely special case where each site is occupied by a boson.
Any single translation for all particles will return to itself (up to a phase factor), and its cyclic subspace has only one element, which is itself.
The only permissible momentum it can form is $K_x=K_y=0$.
We can therefore learn that, for certain momenta, some seed bases are not applicable to construct the momentum basis, leading to the momentum selection rule.
Before using seed basis to construct momentum basis, we need to specify which momentum value is valid.
There is a convenient method to determine if a given momentum value is commensurate to a specific seed basis.
It is known that, by simply translating the seed basis along $x$ and $y$ directions for $L_x$ and $L_y$ times, the seed basis must return to itself. 
If a momentum value is not commensurate for this seed basis, the summation in Eq.~\ref{eqn:K_basis_direct} will produce a zero result \cite{Sandvik2010, jung2020guide}.
%
This offers an efficient approach to identifying the permissible momentum values and can be implemented efficiently in numerical computation.
However, it is not straightforward to understand why certain momentum values are inaccessible to specific seed states. 
To address this, we introduce a systematic method in what follows to explicitly identify permissible momenta for each seed basis and subsequently construct the corresponding momentum basis in real space and explore the reasons behind it.

\subsection{Constructing momentum basis along one direction}
Firstly, we start with the seed basis ${|\rm{seed},\beta \rangle }$ and apply the translation operator $\hat{T}_\mu, \mu=x,y$ repeatedly until we traverse its cyclic subspace.
Each seed basis generates a cyclic subspace under the action of the translation group:
\begin{equation}
\left\{ {\left. {{{\left( {{{\hat T}_\mu }} \right)}^m}|{\rm{seed}},\beta \rangle } \right|m = 0, \cdots ,{{\cal N}^\mu_\beta } - 1} \right\},
\end{equation}
in which $\mathcal{N}^\mu_\beta$ denotes the number of this set.
This procedure can be implemented efficiently in numerical computation.
For single-particle system ($N=1$), there must be $\mathcal{N}^{\mu}_\beta = L_\mu$, since we need to apply the translation operator for $L_\mu$ times to return the initial state.
However, for multi-particle systems ($N>1$), the dimension of this subspace $\mathcal{N}^\mu_\beta$ may vary for different seed bases.
In some peculiar cases, we only need to translate $\mathcal{N}^{\mu}_\beta < L_\mu$ times to return to the original seed basis.
This is due to the indistinguishable nature of multi-particle quantum system under periodic boundary condition in finite system, as mentioned in Ref.~\cite{PhysRevB.107.125161}.
This also imposes a limitation on the permissible values of the momentum for certain seed basis.
In addition, when translating back to the seed basis and $\mathcal{N}^{\mu}_\beta  < L_\mu$, there can be a phase factor:
\begin{equation}
\label{eqn:translation_phase_factor_single_direction}
(\hat{T}_\mu) ^{\mathcal{N}^{\mu}_\beta } |{\rm{seed}},\beta \rangle = {e^{i\Theta \left( \mu, \beta  \right)}}|{\rm{seed}},\beta \rangle 
\end{equation}
due to the particle statistics.
This is somewhat similar to imposing a twisted boundary condition to this cyclic translation \cite{Niu_1984, PhysRevB.103.224208, PhysRevB.107.125161, PhysRevB.108.174204}.
For convenience, we shall call $\Theta \left( \mu, \beta  \right)$ the \emph{translation phase factor}.
Then, the corresponding eigenstate of the translation operator along $\mu=x,y$ direction can be found to possess a following plane-wave-like form:
\begin{equation}
\label{eqn:K_basis_single_direction}
|{{K}_\mu },\beta \rangle  = \frac{1}{{\sqrt {{\cal N}_\beta ^\mu } }}\sum\limits_{m = 0}^{{\cal N}_\beta ^\mu  - 1} {{e^{i\left( {{{\kappa}_\mu } - \frac{{\Theta \left( {\mu ,\beta } \right)}}{{{\cal N}_\beta ^\mu }}} \right)m}}{{\left( {{{\hat T}_\mu }} \right)}^m}|{\rm{seed}},\beta \rangle }  ,
\end{equation}
where ${\kappa}_\mu = 2n_{\mu}\pi/{{\cal N}_\beta ^\mu }$ is the ``bare" momentum with $ n_{\mu} = 0,\cdots,{\cal N}_\beta ^\mu - 1$.
It can be verified that this is the eigenstate of translation operator $\hat{T}_\mu$:
\begin{eqnarray}
{{\hat T}_\mu }|{{K}_\mu },\beta \rangle  = {e^{ - i {K_\mu }} }|{{K}_\mu },\beta \rangle ,
\end{eqnarray}
with a ``renormalized" momentum ${{ K}_\mu } = {{\kappa}_\mu } - \frac{{\Theta \left( {\mu ,\beta } \right)}}{{{\cal N}_\beta ^\mu }}$.
For bosonic systems, the phase factor $\Theta \left( \mu, \beta  \right)$ simply becomes zero.
Otherwise, this phase shift could be non-zero, and should be carefully evaluated in numerical computation.
In particular, the extra phase also poses restriction to permissible momentum values for the momentum state based on this seed basis.
So far, we have only discussed the construction of eigenstates along single direction.
The discussion above can be readily applied to the 1D system.

\subsubsection*{Example for a non-trivial case in 1D systems}
Here we demonstrate a non-trivial case with non-zero phase factor $\Theta \left( \mu, \beta  \right)$ as a typical example.
For convenience, we consider a simple 1D lattice with $L=4$ sites and $N=2$ fermions under PBC.
The seed basis can be easily found:
\begin{eqnarray}
|{\rm{Seed}},{\beta _1}\rangle  &=&  \hat c_1^\dag \hat c_2^\dag |0\rangle ;\nonumber  \\
|{\rm{Seed}},{\beta _2}\rangle  &=&  \hat c_1^\dag \hat c_3^\dag |0\rangle .
\end{eqnarray}
By applying the translation operator $\hat T\hat c_j^\dag {{\hat T}^{ - 1}} = \hat c_{j + 1}^\dag $ successively to these two states, we have following relations
\begin{eqnarray}
{{\hat T}^4}|{\rm{Seed}},{\beta _1}\rangle  = {{\hat T}^4}\hat c_1^\dag \hat c_2^\dag {\left( {{{\hat T}^{ - 1}}\hat T} \right)^4}|0\rangle \nonumber \\
 = \hat c_5^\dag \hat c_6^\dag |0\rangle  = \hat c_1^\dag \hat c_2^\dag |0\rangle  = |{\rm{Seed}},{\beta _1}\rangle 
\end{eqnarray}
and
\begin{eqnarray}
{{\hat T}^2}|{\rm{Seed}},{\beta _2}\rangle  &=& {{\hat T}^2}\hat c_1^\dag \hat c_3^\dag {\left( {{{\hat T}^{ - 1}}\hat T} \right)^2}|0\rangle \nonumber \\
 &=& \hat c_3^\dag \hat c_5^\dag |0\rangle  = \hat c_3^\dag \hat c_1^\dag |0\rangle  =  - \hat c_1^\dag \hat c_3^\dag |0\rangle\nonumber   \\ 
 &=&  e^{i\pi} |{\rm{Seed}},{\beta _2}\rangle ,
\end{eqnarray}
in which we have used the fermionic commutation relation $\hat c_i^\dag \hat c_j^\dag  =  - \hat c_j^\dag \hat c_i^\dag ,\;i \ne j$ and the periodic boundary relation $\hat c_{j + L}^\dag  = \hat c_j^\dag $.
It can be clearly seen that the dimension of the cyclic subspace generated from $|{\rm{Seed}},{\beta _1}\rangle$ is $\mathcal{N}_{\beta_1} = 4$, while it is $\mathcal{N}_{\beta_2} = 2$ for $|{\rm{Seed}},{\beta _2}\rangle$.
The momentum basis for $|{\rm{Seed}},{\beta _1}\rangle$ can be therefore written as
\begin{eqnarray}
|K,{\beta _1}\rangle  = \frac{1}{{\sqrt 4 }}\sum\limits_{m = 0}^3 {{e^{i\kappa m}}{{\hat T}^m}|{\rm{Seed}},{\beta _1}\rangle } ,
\end{eqnarray}
with $\kappa  = \frac{{2n\pi }}{{{{\cal N}_{{\beta _1}}}}} = \frac{{n\pi }}{2},n = 0,1,2,3$ and $K=\kappa$.
As for $|{\rm{Seed}},{\beta _2}\rangle$, applying $\hat{T}^2$ leads to a sign change due to  fermionic statistics, resulting in a translation phase factor $\Theta \left( \beta_2  \right) =\pi$ in Eq.~\ref{eqn:translation_phase_factor_single_direction}.
According to Eq.~\ref{eqn:K_basis_single_direction}, the momentum basis for $|{\rm{Seed}},{\beta _2}\rangle$ reads as
\begin{eqnarray}
|K,{\beta _2}\rangle  = \frac{1}{{\sqrt 2 }}\sum\limits_{m = 0}^1 {{e^{i\left( {\kappa  + \frac{\pi }{2}} \right)m}}{{\hat T}^m}|{\rm{Seed}},{\beta _2}\rangle } ,
\end{eqnarray}
where the bare momentum is $\kappa  = \frac{{2n\pi }}{{{{\cal N}_{{\beta _2}}}}} = n\pi ,n = 0,1$.
The renormalized momentum is given by $K = \kappa  + \frac{\pi }{2}$.
The possible momenta of $|K,{\beta _2}\rangle $ is restricted to $K=\pi/2$ and $K=3\pi/2$.
It can be verified that
\begin{eqnarray}
\hat T|K,{\beta _2}\rangle  &=&  \frac{1}{{\sqrt 2 }}\sum\limits_{m = 0}^1 {{e^{i\left( {\kappa  + \frac{\pi }{2}} \right)m}}{{\hat T}^{m + 1}}|{\rm{Seed}},{\beta _2}\rangle } ,  \nonumber \\
&=&  \frac{1}{{\sqrt 2 }}{\left( {\hat T + {e^{i\left( {\kappa  + \frac{\pi }{2}} \right)}}{{\hat T}^2}} \right)|{\rm{Seed}},{\beta _2}\rangle }  \nonumber\\
&=&  \frac{1}{{\sqrt 2 }}{\left( {\hat T + {e^{i\left( {\kappa  + \frac{\pi }{2}} \right)}}{e^{i\pi }}} \right)|{\rm{Seed}},{\beta _2}\rangle }  \nonumber \\
&=&  {e^{ - i\left( {\kappa  + \frac{\pi }{2}} \right)}}\frac{1}{{\sqrt 2 }} {\left( {\hat T{e^{i\left( {\kappa  + \frac{\pi }{2}} \right)}} + {e^{i2\kappa }}} \right)|{\rm{Seed}},{\beta _2}\rangle }  \nonumber \\
&=&  {e^{ - i\left( {\kappa  + \frac{\pi }{2}} \right)}}|K,{\beta _2}\rangle ,
\end{eqnarray}
where we use the fact that ${e^{i2\kappa }} = {e^{i2n\pi }} = 1, n=0,1$.
We would like to emphasize that this is purely the nature of the multi-particle system.

\subsection{Constructing momentum basis along the other direction}
Next, we construct the simultaneous eigenstate of $\hat{T}_x$ and $\hat{T}_y$ using a similar method .
Suppose that we have constructed the momentum basis along $x$ direction using Eq.~\ref{eqn:K_basis_single_direction}, we then use them to construct the eigenstate of $\hat{T}_y$ since $\left[ {{{\hat T}_x},{{\hat T}_y}} \right] = 0$.
By applying $\hat{T}_y$ to arbitrary $K_x$ momentum basis, it will generate a set of states and finally return to the initial state.
Similarly, it forms a cyclic subspace of the translation operator $\hat{T}_y$:
\begin{equation}
\left\{ {\left. {{{\left( {{{\hat T}_y}} \right)}^m}|{K_x},\beta_{\rm{seed}} \rangle } \right|m = 0, \cdots ,{\cal N}_\beta ^{x,y} - 1} \right\},
\end{equation}
where ${\cal N}_\beta ^{x,y}$ denotes the total number of states in this cyclic subspace and $|{K_x},\beta_{\rm{seed}} \rangle$ is denoted as the seed basis (cyclic vector).
Through this method, we can find all cyclic subspaces of $\hat{T}_y$ from the $K_x$ momentum basis.
Next, we can construct the eigenstate of $\hat{T}_y$ based on these states.
From the cyclic relation, we expect the eigenstate of $\hat{T}_y$ should also have a plane-wave form, like Eq.~\ref{eqn:K_basis_single_direction}.
Similar to the previous discussion in 1D, we find that the cyclic translation may lead to a phase:
\begin{equation}
\label{eqn:translation_phase_factor}
{\left( {{{\hat T}_y}} \right)^{{\cal N}_\beta ^{x,y}}}|{K_x },{\beta _{{\rm{seed}}}}\rangle  = {e^{i\Phi \left( {{K_x },{\beta _{{\rm{seed}}}}} \right)}}|{K_x },{\beta _{{\rm{seed}}}}\rangle ,
\end{equation}
in which the ${\Phi \left( {{K_x },{\beta _{{\rm{seed}}}}} \right)}$ also appears as a translation phase factor.
Remarkably, the appearance of this phase factor is not only related to the particle statistics, but also due to the multi-particle nature.
This is in stark contrast to the 1D case and we will illustrate this point with an example in the following context.
Hence, with some similar calculations, the eigenstate of $\hat{T}_y$ can be solved analytically:
\begin{equation}
\label{eqn:K_basis_two_direction}
|{K_x},{K_y},\beta \rangle  = \frac{1}{{\sqrt {{\cal N}_\beta ^{x,y}} }}\sum\limits_{m = 0}^{{\cal N}_\beta ^{x,y} - 1} {{e^{i{K_y}m}}{{\left( {{{\hat T}_y}} \right)}^m}|{K_x},{\beta _{{\rm{seed}}}}\rangle },
\end{equation}
with
\begin{equation}
\label{eqn:Ky_related_to_Kx}
K_y = {{{\kappa}_y} - \frac{{\Phi \left( {{K_x},{\beta _{{\rm{seed}}}}} \right)}}{{{\cal N}_\beta ^{x,y}}}}
\end{equation}
being the renormalized momentum, which is related to $K_x$.
It can be seen that the solution is still a plane wave up to a phase shift.
Combining with Eq.~\ref{eqn:K_basis_single_direction}, we have
\begin{eqnarray}
\label{eqn:K_basis_two_direction_expanded}
|{\bf{K}},\beta \rangle  &=& \frac{1}{{\sqrt {{\cal N}_\beta ^x{\cal N}_\beta ^{x,y}} }}\sum\limits_{m = 0}^{{\cal N}_\beta ^x - 1} {\sum\limits_{l = 0}^{{\cal N}_\beta ^{x,y} - 1} {{e^{i{K_y}l}}{e^{i{K_x}m}} } } \nonumber \\
&& \qquad \qquad  \times {{\left( {{{\hat T}_y}} \right)}^l}{{\left( {{{\hat T}_x}} \right)}^m}|{\rm{seed}},\beta \rangle ,
\end{eqnarray}
which is quite similar to the single-particle case except for renormalized momenta due to many-body effects under PBC.
It can be seen that, in 1D, the permissible value of momentum $K$ for a seed basis only depends on dimensions of the corresponding cyclic subspace. 
In 2D, however, it can be seen from Eq.~\ref{eqn:Ky_related_to_Kx} that the possible values of $K_y$ also depends on $K_x$, which should be carefully treated in numerical computation.
This reveals the non-trivial feature in higher dimensions.
For real-space seed states possessing high spatial symmetry, translational operations in position space inevitably induce cross-constraints.
This implies that the simultaneous eigenstates of the operators $\hat{T}_x$ and $\hat{T}_y$ must satisfy specific consistency conditions.
This constraint invalidates the conventional assumption that $K_x$ and $K_y$ function as independent degrees of freedom, resulting in an algebraic entanglement of momentum quantum numbers in orthogonal directions within specific high-symmetry subspaces.
Consequently, the allowed momentum values $(K_x, K_y)$ are governed by strict selection rules, where the momentum eigenvalue in one direction analytically depends on the momentum value in its orthogonal direction.
Mathematically, the emergence of these additional phases does not imply a violation of the commutativity between the global translation operators $\hat{T}_x$ and $\hat{T}_y$. 
Instead, they stem from the non-trivial relations within the stabilizer subgroup (or little group) of the specific seed configuration orbit. 
These phase factors act as one-dimensional characters of the stabilizer subgroup, enforcing strict compatibility conditions between the spatial symmetries of the many-body configuration and the Bloch phases. 
Consequently, the symmetry of the real-space configuration reduces the independent degrees of freedom of the center-of-mass momentum, shifting and constraining the allowed $(K_x, K_y)$ eigenvalue pairs through these geometric stabilizer constraints.

\subsubsection*{General proof of the momentum coupling condition}
We now provide a simple and general proof of this momentum coupling condition.
For a given seed basis $|{\rm{seed}},\beta\rangle$, consider all the translations that leave it invariant up to a phase,
\begin{equation}
\label{eqn:stabilizer_def}
{{\hat T}_x}^{a}{{\hat T}_y}^{b}|{\rm{seed}},\beta\rangle = {e^{i\theta_{a,b}}}|{\rm{seed}},\beta\rangle .
\end{equation}
Such translations form the stabilizer subgroup $S_\beta$ of the seed basis, and since $\left[ {{{\hat T}_x},{{\hat T}_y}} \right] = 0$, the corresponding phases satisfy $\theta_{a+a',b+b'}=\theta_{a,b}+\theta_{a',b'}$ (mod $2\pi$), i.e., they act as a one-dimensional character of $S_\beta$.
Because the momentum basis $|{K_x},{K_y},\beta \rangle$ is a simultaneous eigenstate of ${{\hat T}_x}$ and ${{\hat T}_y}$ with eigenvalues ${e^{-i{K_x}}}$ and ${e^{-i{K_y}}}$, acting with Eq.~\ref{eqn:stabilizer_def} on it immediately yields the phase-matching condition
\begin{equation}
\label{eqn:phase_matching}
a{K_x} + b{K_y} + \theta_{a,b} \equiv 0 \pmod{2\pi},
\end{equation}
which must hold for every element $(a,b)\in S_\beta$.
Conversely, if Eq.~\ref{eqn:phase_matching} is violated for some stabilizer element, the Bloch phases in the plane-wave superposition of Eq.~\ref{eqn:K_basis_direct} sum to zero and the constructed state vanishes identically.
This is the group-theoretical origin of the zero-result criterion discussed below Eq.~\ref{eqn:K_basis_direct}.
Therefore, the allowed momentum pairs of a seed basis are exactly the solutions of Eq.~\ref{eqn:phase_matching} over its stabilizer subgroup.
Eq.~\ref{eqn:phase_matching} also clarifies when $K_x$ and $K_y$ can be chosen independently.
If $S_\beta$ is generated by pure translations along the two directions, i.e., elements of the form $(a,0)$ and $(0,b)$, then every constraint involves either $K_x$ or $K_y$ alone, and the two momenta are decoupled.
By contrast, if $S_\beta$ contains a mixed translation $(a,b)$ with $a\neq 0$ and $b\neq 0$ that cannot be generated by pure translations, the corresponding constraint in Eq.~\ref{eqn:phase_matching} links $K_x$ and $K_y$ algebraically.
This usually happens when the particle distribution possesses a diagonal translational symmetry.
The cross-dimensional momentum coupling is thus a general consequence of such diagonal symmetry, rather than a peculiarity of any specific example.
Moreover, if the lattice and the seed configuration are invariant under the exchange $x\leftrightarrow y$ up to a phase, then $(a,b)\in S_\beta$ implies $(b,a)\in S_\beta$ with $\theta_{b,a}=\theta_{a,b}$ (mod $2\pi$), so that the constraint set of Eq.~\ref{eqn:phase_matching} is invariant under $K_x\leftrightarrow K_y$.
This explains in general why the allowed momentum values are symmetric about $K_x=K_y$, as observed in Table~\ref{table:possib_Ky}.

Finally, Eq.~\ref{eqn:phase_matching} underlies the sequential construction in Eqs.~\ref{eqn:translation_phase_factor}--\ref{eqn:Ky_related_to_Kx}.
Let $(a_0,b_0)$ be a stabilizer element with the smallest positive $b_0$; then the cyclic subspace of ${{\hat T}_y}$ generated from $|{K_x},{\beta _{{\rm{seed}}}}\rangle$ closes after $b_0$ steps, ${\cal N}_\beta ^{x,y}=b_0$, and solving Eq.~\ref{eqn:phase_matching} for $K_y$ gives
\begin{equation}
\label{eqn:Ky_general}
{K_y}=\frac{2\pi n_y}{b_0}-\frac{\theta_{a_0,b_0}+a_0K_x}{b_0},\qquad n_y=0,\cdots ,b_0-1,
\end{equation}
which is exactly Eq.~\ref{eqn:Ky_related_to_Kx} with the translation phase factor
\begin{equation}
\label{eqn:Phi_general}
\Phi \left( {{K_x},{\beta _{{\rm{seed}}}}} \right) = \theta_{a_0,b_0}+a_0K_x .
\end{equation}
The result is independent of the choice of $a_0$ modulo $2\pi$, since different choices differ by a pure-$x$ translation whose constraint has already been imposed on $K_x$.
For the $4\times4$ example discussed below, the seed basis satisfies ${{\hat T}_y}^2|{\rm{seed}},\beta \rangle={{\hat T}_x}^2|{\rm{seed}},\beta \rangle$, corresponding to $(a_0,b_0)=(-2,2)$ with $\theta_{-2,2}=0$; Eq.~\ref{eqn:Phi_general} then yields $\Phi=-2K_x$, and Eq.~\ref{eqn:Ky_general} gives $K_y=\pi n_y+K_x$, in full agreement with the direct calculation and Table~\ref{table:possib_Ky}.

\subsubsection*{Example for a non-trivial case in 2D systems}
Previously, we mentioned that the non-zero translation phase factor ${\Phi \left( {{K_\mu },{\beta _{{\rm{seed}}}}} \right)}$ arising in constructing the momentum basis $|{\bf{K}},\beta \rangle$ can result from multi-particle symmetry.
This phenomenon is subtle, and we will adopt a $4\times 4$ system with $N=2$ bosons under PBC as a typical example to elucidate this point.
For bosonic systems, the particle statistics effects on translation phase factors ($\Theta \left( \mu, \beta  \right)$ and ${\Phi \left( {{K_\mu },{\beta _{{\rm{seed}}}}} \right)}$) are both suppressed.
Under this condition, we find a special momentum basis with non-zero translation phase factor ${\Phi \left( {{K_\mu },{\beta _{{\rm{seed}}}}} \right)} \ne 0$.
The corresponding seed basis is found to be $|{\rm{seed}},\beta \rangle  = |{{\bf{r}}_1},{{\bf{r}}_2}\rangle $ with ${{\bf{r}}_1} = \left( {1,1} \right);{{\bf{r}}_2} = \left( {3,3} \right) $.
To facilitate a clearer understanding, we express the seed basis as
\begin{equation}
|{\rm{seed}},\beta \rangle  = | {\vcenter{\hbox{\includegraphics[height = 0.75cm]{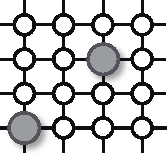}}} }  \rangle.
\end{equation}
In $x$ direction, there is $\hat{T}_x^{\mathcal{N}_\beta ^x}|{\rm{seed}},\beta \rangle = |{\rm{seed}},\beta \rangle$ with ${\mathcal{N}_\beta ^x} = 4$.
From Eq.~\ref{eqn:translation_phase_factor_single_direction}, one can immediately find the translation factor: $\Theta(x,\beta) = 0$.
Therefore, we can construct the eigenstate of $\hat{T}_x$ according to Eq.~\ref{eqn:K_basis_single_direction}:
\begin{equation}
|{K_x},\beta_{\rm{seed}} \rangle  = \frac{1}{{\sqrt 4 }}\sum\limits_{m = 0}^3 { {e^{i{K_x}m}}{\hat T_x}^m  | {\vcenter{\hbox{\includegraphics[height = 0.75cm]{EQ_SeedBasis.eps}}} }  \rangle } ,
\end{equation}
with $K_x = 2n\pi/4, n=0,1,2,3$.
Next, we use these states to construct the eigenstates of $\hat{T}_y$.
By applying the translation operator along $y$ direction, we find
\begin{equation}
\hat{T}_y|{K_x},\beta_{\rm{seed}} \rangle  = \frac{1}{{\sqrt 4 }}\sum\limits_{m = 0}^3 { {e^{i{K_x}m}}{\hat T_x}^m  | {\vcenter{\hbox{\includegraphics[height = 0.75cm]{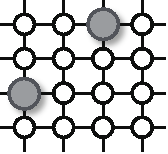}}} }  \rangle } , 
\end{equation}
and
\begin{eqnarray}
\hat{T}_y^2|{K_x},\beta_{\rm{seed}} \rangle  &=& \frac{1}{{\sqrt 4 }}\sum\limits_{m = 0}^3 { {e^{i{K_x}m}}{\hat T_x}^m  | {\vcenter{\hbox{\includegraphics[height = 0.75cm]{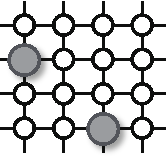}}} }  \rangle }  \nonumber \\
&=& \frac{1}{{\sqrt 4 }}\sum\limits_{m = 0}^3 { {e^{i{K_x}m}}{\hat T_x}^{m+2}  | {\vcenter{\hbox{\includegraphics[height = 0.75cm]{EQ_SeedBasis.eps}}} }  \rangle }  \nonumber \\
&=& e^{-2iK_x} |{K_x},\beta_{\rm{seed}} \rangle  .
\end{eqnarray}
It can be seen that the dimension of the cyclic subspace of $\hat{T}_y$ generated by $|{K_x},\beta_{\rm{seed}} \rangle $ is $\mathcal{N}_\beta^{x,y} = 2$, which is smaller than the system's size $L_y=4$.
According to Eq.~\ref{eqn:translation_phase_factor}, one can compute a non-zero translation phase factor:
\begin{equation}
\Phi \left( {{K_x },{\beta _{{\rm{seed}}}}} \right) = -2K_x.
\end{equation}
Thus, according to Eq.~\ref{eqn:Ky_related_to_Kx} and Eq.~\ref{eqn:K_basis_two_direction_expanded}, the momentum basis reads as
\begin{eqnarray}
\label{eqn:example_KxKy_4x4}
|{\bf{K}},\beta \rangle  &=& \frac{1}{{\sqrt 8 }}\sum\limits_{m = 0}^3 {\sum\limits_{l = 0}^1 {{e^{i{K_y}l}}{e^{i{K_x}m}}} } \nonumber \\
&& \qquad \times {{\left( {{{\hat T}_y}} \right)}^l}{{\left( {{{\hat T}_x}} \right)}^m}   | {\vcenter{\hbox{\includegraphics[height = 0.75cm]{EQ_SeedBasis.eps}}} }  \rangle.
\end{eqnarray}
with $K_x =\frac{ 2n_x \pi}{4}, n_x=0,1,2,3$ and renormalized momentum $K_y = \frac{2n_y \pi}{2}  + {K_x}, n_y=0,1$.
It can be found in this specific case that $K_y$ is closely related to the value of $K_x$.
This restricts the possible values of $(K_x, K_y)$ pairs for certain seed basis, see Table~\ref{table:possib_Ky} for example.
Hence, one should carefully determine the values of $K_x$ and $K_y$ in practical calculations.
Notably, if one starts by determining $K_x$ and then $K_y$, the relation is similar.
We would like to emphasize that this result is extremely particular.
Since translation operators along $x$ and $y$ are independent to each other, their eigenstate should also be independent and $K_x$ should not be coupled to $K_y$.
The phenomenon described above can be attributed to the special symmetry of this seed basis state under diagonal translations (the combination of $x$ and $y$ translation).

\begin{table}[!ht]
    \centering
    \begin{tabular}{{|c|c|c|c|c|}}
    \hline
        $K_x$ & 0 & $\pi/2$ &$ \pi$ &$ 3\pi/2$ \\ \hline
       $ K_y$ & $0, \pi$ & $\pi/2, 3\pi/2$ & $0,\pi$ & $\pi/2, 3\pi/2$ \\ \hline
    \end{tabular}
    \caption{
    \label{table:possib_Ky}
Possible values of $K_y$ for different $K_x$ in Eq.~\ref{eqn:example_KxKy_4x4}.
}
\end{table}

This conditional dependence of $K_y$ on $K_x$ is of vital importance for developing practical ED codes. 
In standard single-particle or non-symmetric multi-particle systems, loops over $K_x$ and $K_y$ are typically nested independently. 
However, for the high-symmetry configurations discovered here, treating $K_x$ and $K_y$ as independent degrees of freedom would cause the ED algorithm to either generate non-existent (forbidden) states or repeatedly count identical state vectors across different momentum sectors, destroying the hermiticity and correctness of the block-diagonalized Hamiltonian.

\subsection{Constructing Bloch Hamiltonian in real space}
Eq.~\ref{eqn:K_basis_two_direction_expanded} provides a guide to construct momentum basis and the Bloch Hamiltonian in numerical computation.
Here, we briefly demonstrate the method, which has been also covered in Ref.~\cite{jung2020guide}.
According to Eq.~\ref{eqn:K_basis_two_direction_expanded}, the matrix element of the Bloch Hamiltonian in the $(K_x, K_y)$ subspace reads as:
\begin{eqnarray}
\label{eqn:Bloch_Ham_matrix_element}
{\left[ {H\left( {\bf{K}} \right)} \right]_{\beta ,\beta '}} &=& \langle {\bf{K}},\beta |\hat H|{\bf{K}},\beta' \rangle \nonumber \\
&=&\frac{{\sum\limits_{m = 0}^{N_\beta ^x - 1} {\sum\limits_{l = 0}^{N_\beta ^{x,y} - 1} {\sum\limits_{m' = 0}^{N_{\beta '}^x - 1} {\sum\limits_{l' = 0}^{N_{\beta '}^{x,y} - 1} {{e^{i{K_y}\left( {l' - l} \right)}}{e^{i{K_x}\left( {m' - m} \right)}}} \langle {\rm{seed}},\beta |{{\hat T}_y}^{l' - l}{{\hat T}_x}^{m' - m}\hat H|{\rm{seed}},\beta '\rangle } } } }}{{\sqrt {N_\beta ^xN_\beta ^{x,y}} \sqrt {N_{\beta '}^xN_{\beta '}^{x,y}} }}
\end{eqnarray}
where we have used the commutation relation: $[ {{{\hat T}_\mu },\hat H} ] = 0$.
It means that, without constructing the full Hamiltonian matrix, we can only focus on seed bases and evaluate the matrix element of Bloch Hamiltonian $H(\boldsymbol{K})$ with the help of translation operators in real space.
Moreover, it can be noted that the matrix element is non-zero only if $\hat H|{\rm{seed}},\beta '\rangle  \propto {{\hat T}_y}^{\Delta l}{{\hat T}_x}^{\Delta m}|{\rm{seed}},\beta \rangle $.
As long as the relative shifts $\Delta l$ and $\Delta m $ are determined, the summations in Eq.~\ref{eqn:Bloch_Ham_matrix_element} will be greatly simplified.
In numerical computation, it is extremely beneficial that we calculate and store all possible relative distances $\left\{ {\Delta l,\Delta m} \right\}$, the mapping relation of matrix element $\left\{ {\beta,\beta'} \right\}$, normalization factors ${\cal N}_\beta ^x{\cal N}_\beta ^{x,y}$, and tunneling matrix elements in advance by defining and computing:
\begin{equation}
M\left( {\beta ,\beta ',\Delta m,\Delta l} \right) \equiv \frac{{\langle {\rm{seed}},\beta |{{\hat T}_y}^{\Delta l}{{\hat T}_x}^{\Delta m}\hat H|{\rm{seed}},\beta '\rangle }}{{\sqrt {{\cal N}_\beta ^x{\cal N}_\beta ^{x,y}} \sqrt {{\cal N}_{\beta '}^x{\cal N}_{\beta '}^{x,y}} }}.
\end{equation}
Notably, if the particle is not bosonic, one needs further store the permutation phase due to the particle statistics and we only consider the bosons here for simplicity.
The matrix element can be therefore greatly simplified as
\begin{equation}
{\left[ {H\left( {\bf{K}} \right)} \right]_{\beta ,\beta '}} = \sum\limits_{\Delta m,\Delta l}^{} {{e^{i{K_y}\Delta l}}{e^{i{K_x}\Delta m}}M\left( {\beta ,\beta ',\Delta m,\Delta l} \right)} 
\end{equation}
Then, for each momentum value $\boldsymbol{K}$, one can directly use these real-space matrix elements without repeated calculations to improve the efficiency of the numerical calculation.
On the other hand, different from the momentum-space approach, the interaction term is now diagonal in the Bloch Hamiltonian, while tunneling terms are usually off-diagonal.
We note that the matrix element of the tunneling terms are related to the range of tunneling, which is usually finite.
Hence, the Bloch Hamiltonian can be very sparse, which is beneficial for saving the computation resource.
The above procedure can be efficiently implemented with the exact diagonalization algorithm introduced in Ref.~\cite{Zhang_2010}.

\section{Numerical computation}
\label{sec:numerical_computation}
\begin{figure}
  \includegraphics[width = 0.9\columnwidth ]{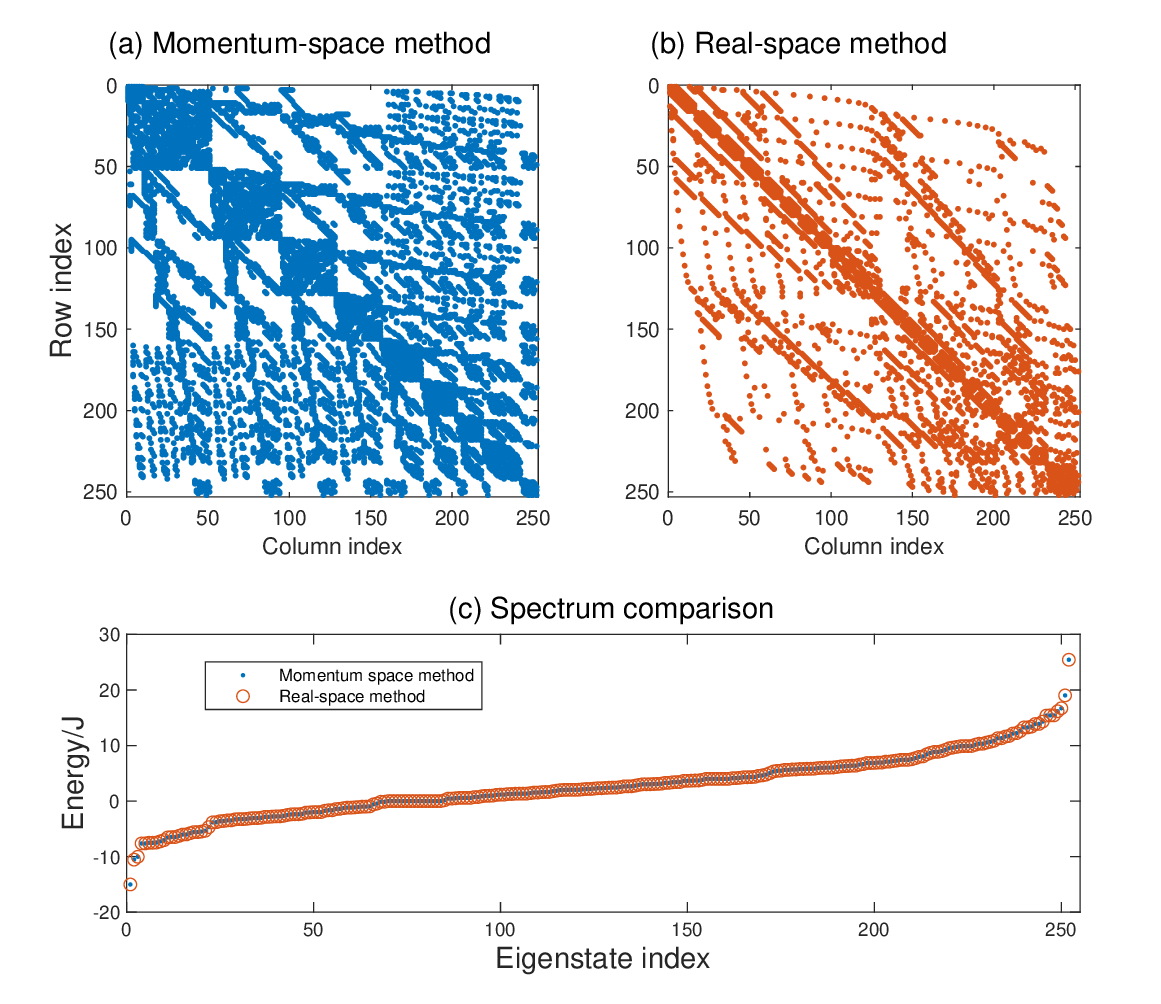}
  \caption{\label{fig:Figure_Hk_nonzeros}
 Exact diagonalization for the Bose-Hubbard model with total momentum $K_x=K_y=0$ in a $L_x = L_y = 4$ square lattice with total $N=4$ bosons.
 (a-b) Non-zero matrix elements of c.m. Bloch Hamiltonian computed from two different methods.
 The number of non-zero matrix element is $M_{nz} = 9228$ in (a) and $M_{nz} = 3298$ in (b).
 (c) Eigenvalues of $H(\mathbf{K})$ computed from the two methods.
 Other parameters are chosen as: $U/J=4$.
  }
\end{figure}

\begin{figure}
  \includegraphics[width = 0.9\columnwidth ]{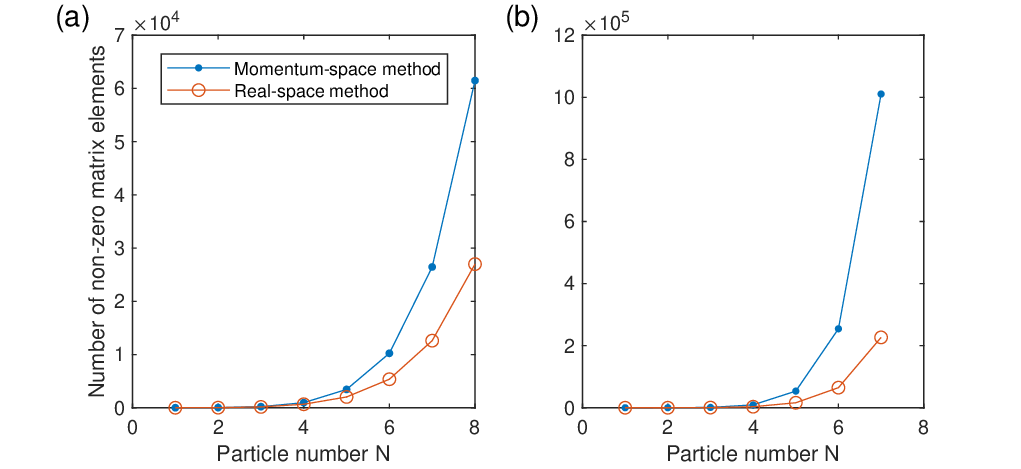}
  \caption{\label{fig:Figure_Nz_element_scan}
 Non-zero matrix element in Bloch Hamiltonian as a function of particle number.
 Here we consider two sizes: (a) $L_x=L_y=3$; and (b) $L_x=L_y=4$.
  }
\end{figure}

In this section, we supplement some simple numerical calculations based on the momentum-space method introduced in Sec.~\ref{sec:Momentum_space_Method} and real-space momentum basis method introduced in Sec.~\ref{sec:Real_space_Method}.
We consider the Bose-Hubbard model as the example, whose Hamiltonian can be written as
\begin{equation}
\hat{H} = -J \sum_{\langle i,j \rangle} (\hat{b}_i^\dag \hat{b}_j + \mathrm{h.c.}) + \frac{U}{2} \sum_i n_i (n_i - 1)
\end{equation}
in real space.
Through the Fourier transformation $ \hat{b}_j = \frac{1}{\sqrt{N_s}} \sum_{\mathbf{k}} e^{i \mathbf{k} \cdot \mathbf{R}_j} \hat{b}_{\mathbf{k}}$, with $N_s$ being the number of total sites, the kinetic part is diagonalized, and the Hamiltonian can be expressed in momentum space:
\begin{equation}
\hat{H} = \sum_{\mathbf{k}} \epsilon_{\mathbf{k}} \hat{b}_{\mathbf{k}}^\dagger \hat{b}_{\mathbf{k}} + \frac{U}{2N_s} \sum_{\mathbf{k}_1, \mathbf{k}_2, \mathbf{k}_3, \mathbf{k}_4} \hat{b}_{\mathbf{k}_1}^\dagger \hat{b}_{\mathbf{k}_2}^\dagger \hat{b}_{\mathbf{k}_3} \hat{b}_{\mathbf{k}_4} \delta_{\mathbf{k}_1+\mathbf{k}_2, \mathbf{k}_3+\mathbf{k}_4 \pmod{\mathbf{G}}}
\end{equation}
in momentum space with $\epsilon_{\mathbf{k}} = -2J(\cos k_x + \cos k_y)$.
Notably, the total momentum is conserved here.
Next, we implement the momentum-space method and the real-space method to perform exact diagonalization calculation, see Fig.~\ref{fig:Figure_Hk_nonzeros}.
It can be seen that, in this relatively small system, the non-zero matrix element from the real-space method is smaller than the momentum-space method.
Meanwhile, the eigenvalues of these two Bloch Hamiltonians produce consistent results, showing the equivalence between two methods.
We further examine the non-zero matrix elements for some different cases in Fig.~\ref{fig:Figure_Nz_element_scan}.
As the total particle number grows, the non-zero matrix element of momentum-space method grows much faster than the real-space method.
By using the sparse matrix method, the real-space method will save considerable computing resources.


\section{Summary and discussion}
\label{sec:summary_and_discussion}
In summary, we have developed a comprehensive theoretical and algorithmic framework for constructing center-of-mass momentum eigenstates in 2D many-body quantum lattice systems using a real-space basis approach.
By systematically extending the c.m. momentum projection from 1D to 2D, we reveal that the striking dimensional variations of c.m. momentum cyclic subspaces are fundamentally governed by the real-space geometric symmetries of the many-body configurations (seed bases).
While the basic numerical algorithm has been discussed in previous works \cite{Sandvik2010, jung2020guide, PhysRevB.107.125161}, our approach provides a clear and intuitive physical picture. 
It explains exactly how translation symmetry, lattice geometry, and quantum statistics work together at the basis level.

Crucially, we uncover that the translation phase factors, arising either from intrinsic quantum statistics or from highly symmetric spatial configurations, can be rigorously formulated as the characters of the stabilizer subgroup associated with each many-body orbit. 
This mechanism imposes effective twisted boundary conditions that inherently renormalize the permissible momentum eigenvalues. 
An interesting discovery of our work is the emergence of an anomalous cross-dimensional momentum coupling in 2D systems: for specific high-symmetry multi-particle configurations, the allowed momentum eigenvalues in orthogonal spatial directions become algebraically entangled. 
Consequently, the momentum components $K_x$ and $K_y$ cease to be independent degrees of freedom, giving rise to emergent momentum selection rules.
We have provided concrete examples to illustrate this feature.
On a broader note, these orbit-dependent momentum selection rules may provide a useful finite-size language for analyzing symmetry constraints related to Lieb-Schultz-Mattis (LSM)-type arguments~\cite{LIEB1961407, PhysRevLett.79.1110, PhysRevLett.84.1535, PhysRevLett.78.1984} and their higher-dimensional generalizations. 
While conventional LSM or Oshikawa-type theorems operate on the thermodynamic limit to restrict ground-state degeneracies and excitation gaps via flux insertion, our real-space formalism exposes how translation symmetries group-theoretically constrain individual multi-particle orbits within a finite lattice. 
Crucially, because flux insertion is mathematically equivalent to shifting the center-of-mass momentum via twisted boundary conditions, our findings also carry direct algorithmic implications for evaluating many-body topological invariants \cite{PhysRevB.107.125161, PhysRevB.108.174204}. 
Specifically, when calculating invariants like the many-body Chern number strictly within isolated high-symmetry sectors, the standard summation over the entire independent $(K_x, K_y)$ Brillouin zone must be restricted to the coupled momentum manifold dictated by our selection rules; otherwise, non-physical zero-states or duplicated basis vectors would lead to ill-defined invariants. 
Although a rigorous and direct connection between our kinematic selection rules and the thermodynamic ground-state constraints requires further investigation, the framework established here offers an intuitive perspective on how spatial symmetries interlock with particle fillings at the un-diagonalized basis level.
Furthermore, we remark on the group-theoretical scope of our current framework. 
The explicit momentum selection rules derived in this work fundamentally rely on the Abelian nature of the pure multi-particle translation group $G = \mathbb{Z}_{L_x} \times \mathbb{Z}_{L_y}$, which guarantees that the non-trivial translation phase factors manifest as one-dimensional characters (i.e., scalar $U(1)$ phases). 
However, if the lattice system possesses additional point-group symmetries (such as $C_4$ rotations or reflections) or operates under a gauge field (magnetic translations), the stabilizer subgroup associated with a highly symmetric many-body configuration may become non-Abelian. 
In such scenarios, our real-space construction framework would naturally generalize by employing higher-dimensional irreducible representations of these non-Abelian groups.
 Consequently, instead of a simple scalar phase shift determining the momentum values, the cyclic operations would induce matrix-valued transformations, leading to the emergence of degenerate multi-particle momentum multiplets. 
Extending the present framework to systematically incorporate non-Abelian spatial and magnetic symmetries represents an intriguing and highly relevant direction for future research.
Our study not only optimizes computational resources for exact diagonalization, but also provides deeper physical insight into the interplay between multi-particle translation symmetry, particle statistics, and many-body quantum physics. 
While our results can be generalized to higher-dimensional systems, future investigations into anyonic system where translation phase factors are expected to be significantly more complex, present an intriguing avenue for research.
Finally, while the specific implications of these highly symmetric c.m. momentum states on ground-state properties remain an open question, the framework established here offers a robust and intuitive pathway for investigating complex quantum phenomena in lattice systems.

\section{Funding}
This work is supported by the Quantum Science and Technology - National Science and Technology Major Project (Grant No. 2025ZD0300800), the National Natural Science Foundation of China (Grants No. 12275365, 92476201, and 12505025), the National Key Research and Development Program of China (Grant No. 2022YFA1404104), the Guangdong Provincial Quantum Science Strategic Initiative (GDZX2305006 and GDZX2405002), and the Natural Science Foundation of Guangdong Province (Grant No. 2023A1515012099).

\vspace*{2mm}

\bibliographystyle{unsrt}

\end{document}